\documentstyle[12pt,aasms4]{article}
\def\simg{\ \lower 3pt\hbox{${\buildrel > \over \sim}$}\ }
\def\siml{\ \lower 3pt\hbox{${\buildrel < \over \sim}$}\ }
\def\reference{\par\noindent\hangindent=1cm\hangafter=1}

\begin{document}
\title{Evidence for early stellar encounters in the
orbital distribution of Edgeworth-Kuiper Belt objects}

\author{Shigeru Ida$^{1,2}$, John Larwood$^{1,3}$, and Andreas Burkert$^{1}$}

\affil{1) Max-Planck-Institut f\"ur Astronomie, K\"onigstuhl 17, D-69117
Heidelberg, Germany}
\affil{2) Dept. of Earth \& Planetary Science, Tokyo Institute of
Technology, Tokyo 152-8551, Japan}
\affil{3) Astronomy Unit, Queen Mary \& Westfield College, London E1 4NS,
United Kingdom}
\authoremail{ida@geo.titech.ac.jp, j.d.larwood@qmw.ac.uk,
burkert@mpia-hd.mpg.de} 

\begin{abstract}
We show that early stellar encounters can explain
the high eccentricities and inclinations observed in the
outer part ($>42$AU) of the Edgeworth-Kuiper Belt (EKB).
We consider the proto-sun as a member of a stellar aggregation
that undergoes dissolution on a timescale $\sim 10^8$ yrs,
such that the solar nebula experiences a flyby encounter
at pericenter distance ($q$) on the order of $100$AU.
Using numerical simulations we show that 
a stellar encounter pumps the velocity dispersion in the young solar nebula
in the outer parts.
In the case of a nearly parabolic encounter with
a solar-mass companion the velocity dispersion at $a \simg 0.25q$ is
pumped up to such an extent
that collisions between planetesimals would be expected to become
highly disruptive, halting further growth of planetesimals.
This has the consequence
that planet formation is forestalled in that region.
We also find that a stellar encounter with pericenter distance 
$q \sim 100$--$200$AU could have pumped up
the velocity dispersion of EKB objects
outside 42AU to the observed magnitude while
preserving that inside Neptune's 3:2
mean-motion resonance (located at 39.5AU). This allows for the
efficient capture of objects by the resonance during a phase of 
orbital migration by proto-Neptune, which we also test with simulations.
We point out that such a stellar encounter generally affects
the dynamical and material structure of a protoplanetary disk and
the planetesimal distribution can remain imprinted with this signature
over much of the main sequence lifetime of the star. In particular,
our results support the notion that an analogous process has
operated in some recently observed extrasolar dust disks.
\end{abstract}


\section{Introduction}

Stars commonly form in groups or clusters within turbulent
molecular clouds on timescales which are of about a million
years (Hillenbrand 1997). Typical young stellar aggregates have sizes
of roughly 1 pc and consist of a few hundred stars. 
Recent observations have also shown that most stars form in
eccentric binary systems and that the binary frequency of young
stars is about two times higher than that of main sequence stars
in the solar neighbourhood (Ghez et al. 1997,
K\"ohler \& Leinert 1998). This reflects the fact that
secular dynamical processes within newly formed stellar groups tend to
reduce their binary fraction over time. Recent numerical
modeling (Kroupa 1995, 1998) demonstrates that encounters between binaries
can lead to the dissolution of aggregates on timescales
of several hundred million years and that stochastic close 
stellar encounters --- which are in general very energetic ---
can lead to the dissociation of the widest binaries. Binary
dissociation
occurs at binary orbital periods greater than about 3000 yrs,
corresponding to separations of order a few 100 AU. 
It is therefore reasonable to
expect that most single main sequence stars actually
formed as part of a wider binary system which was disrupted
through interactions within a young stellar cluster.
Even after a proto-star becomes detached from its companion, or if
it is born as a single star, encounters by passing stars
would occur before the dissolution of the stellar cluster. The
timescale for encounters with pericenter distance $q \sim 200$AU may be
comparable to the dissolution timescale of the stellar cluster
(Laughlin \& Adams 1998). Thus,
if the Sun formed in such a clustered environment,
it most likely experienced a few close encounters with a transient
binary companion or with passing stars at pericenter distances of
order 100 AU, before the break up of the stellar cluster.

Laughlin \& Adams (1998) have suggested that the large eccentricities of 
extrasolar planets associated with Cyg B and 14 Her could have been pumped
up by interactions with passing {\em binary systems} in an open cluster. 
Here, we will consider interactions of a star (the proto-sun)
having a protoplanetary system which encounters a passing single star.
In general interactions with a binary system are more disruptive
to the protoplanetary system than those with a single star.
Since we seek to model the
Solar System the interactions we consider are necessarily much
less disruptive to the planetary system than those considered by
Laughlin \& Adams.
(More distant encounters with passing binary systems may lead to
similar results.)

Such an encounter will generally affect the dynamical and material
structure of
the solar protoplanetary disk and, provided internal conditions allow,
the planetesimal disk will remain imprinted with this signature over
much of the main sequence lifetime of the star.
In this {\it Letter} we study the dynamical effects of the stellar
encounters on protoplanetary disks and point out that 
the orbital distribution of Edgeworth-Kuiper Belt (EKB) objects
may indicate that the Solar System has experienced close
stellar encounters.
We demonstrate that puzzling kinematical features in
the orbital distribution of the EKB objects can be explained
naturally if the Sun formed as a member
of a stellar cluster and experienced a stellar encounter (or series
of encounters) with $q \sim 100$--$200$AU.

\section{Dynamical structure of the Edgeworth-Kuiper Belt}

The observed EKB objects 
observed at multiple oppositions or during relatively long duration
are shown in Fig.1
(e.g. see Marsden's 
web site, http://cfa-www.harvard.edu/$^\sim$graff/lists/TNOs.html). 
The increasing numbers of EKB objects being revealed by observations 
presently fall into three distinct groups.
Firstly, many objects have semimajor axes close to the 3:2
resonance with Neptune's orbit (located at 39.5AU), and these display
a wide range of eccentricities and inclinations (each up to $\sim 0.35$).
Secondly, outside $42$ AU, the objects have slightly lower average 
eccentricity ($\sim 0.1$) and inclination ($\sim 0.1$ radian). 
At semimajor axes inside
$39$AU, and between $40$AU and $42$AU, there are
unpopulated regions (hereinafter "gaps"). 
The cut-off outside $\sim 50$AU may imply depletion of
objects but it could also be due to the present observational sensitivity limit
(Jewitt, Luu, and Trujillo 1998; Gladman et al. 1998).
The third group is comprised of the 'scattered disk' objects
(Duncan and Levison 1997),
which have experienced close approach with Neptune.
Pericenter for the scattered disk objects is located near
Neptune's orbit.
An example is TL66 with $e \sim 0.6$ and $a \sim 85$AU, which is
outside the range of Fig.1.

Secular perturbations by the giant planets can account for 
the gap between $40$AU and $42$AU (Duncan, Levison, \& Budd 1995). They
cannot account for the other features (Duncan, Levison, \& Budd 1995).
The model of sweeping mean motion resonances due to Neptune's outward
migration successfully accounts for the concentrated distribution 
at the 3:2 resonance
as well as for the gap inside $39$AU (Malhotra 1995).
This model also predicts that a large accumulation ought to occur at Neptune's
2:1 resonance (located at $47.8$AU) with a cleared gap interior to
the present resonant location.
If the number of objects captured by the 2:1 sweeping resonance 
is similar to that by the 3:2 resonance,
it may be expected that more objects should now be
detected near the 2:1 resonance (Jewitt, Luu, and Trujillo 1998).
However, the current population near the 2:1 resonance is still
poorly constrained owing to the observational sensitivity limit.
The migration speed of Neptune also affects the relative population
between the 3:2 and 2:1 resonances (Ida et al. 1999). 
In summary, the good agreement of the theoretical predictions by 
Malhotra (1995)
with the observations for the objects near the 3:2 resonance
supports the sweeping of mean motion resonances.

The relatively high eccentricities and inclinations found outside $42$AU
cannot be accounted for by long-range secular perturbations of the planets.
The velocity dispersion of these observed
objects exceeds their surface escape velocity for most objects,
which cannot be explained by internal
gravitational scattering (Safronov 1969).

The capture probability of the sweeping
3:2 resonance becomes small, and the gap inside 39AU
cannot be created, when the initial eccentricity exceeds 
$\sim 0.05$ (Malhotra 1995).
The objects with $e \simg 0.05$ would not be swept and
remain inside 39AU, although a clear gap is presently observed
inside 39AU. Thus, the mechanism to pump up velocity dispersion
outside $42$ AU should satisfy the condition of having 
occurred in a highly localized manner to keep $e$ and $i$
small enough inside 39AU,
although we note that objects with $e \simg 0.1$, inside 39AU,
{\em can} be destabilized by planetary perturbations 
in the age of the Solar System (Duncan, Levison, \& Budd 1995).

Some models have been proposed to account for the high $e$ and $i$ 
outside $42$ AU.
The Earth-sized bodies that are thought to have once existed in the
formation stage and were subsequently ejected
might have been able to pump up the velocity dispersion
(Stern 1991; Morbidelli and Valsecchi 1997; Petit, Morbidelli,
and Valsecchi 1999).
Partial trapping by sweeping of the 2:1 resonance
might have also pumped up the
eccentricities outside $42$AU (Hahn and Malhotra 1999).

Here we propose another mechanism, stellar encounters,
to dynamically heat the planetesimal disk outside $42$AU.
While the two former mechanisms are associated with processes
occurring after the formation of Neptune,
the stellar encounter
model can operate before Neptune's formation as well.
Although all these mechanisms may be able to account for the dynamical 
heating of the velocity dispersion between
the 3:2 and 2:1 resonances, the predicted velocity dispersion
beyond the 2:1 resonance (which has not been observed up to now)
is expected to be quite different in our model, as discussed below.

\section{Modeling}

We have investigated the possibility that stellar encounters 
with the young solar nebula could have increased
the eccentricity $e$ and inclination $i$ of EKB objects presently
located outside $42$AU. 
In our modeling we assume:
\begin{enumerate}
\item A single star passes by the proto-sun 
on a nearly parabolic orbit and perturbs the planetesimal system.
The passing star may be weakly bound to the proto-sun in which case
we can consider a series of encounters.
\item The pericenter distance ($q$) of the encounter(s) is 
on the order of 100AU.
\item Planetesimals with the present EKB object mass
($\sim 10^{22}$--$10^{23}$ g) are formed on
low-$e$ and low-$i$ orbits prior to the first encounter that
pumps up $e$ and $i$ significantly.
\end{enumerate}

As discussed in the introduction, the assumptions 1 and 2 are consistent
with recent observations and numerical modeling.
If we are only concerned with the
effects of stellar encounters on the protoplanetary system,
the assumption 3 is not necessary.
However, to apply our results to the EKB
the assumption 3 is needed in our model because 
the induced velocity dispersion is larger than the surface escape velocity,
which we would expect to halt planetesimal agglomeration (see below).
According to conventional models
(e.g. Safronov 1969; Goldreich \& Ward 1973;
Hayashi, Nakazawa, \& Nakagawa 1985),
dust grains settle to the equatorial plane of the nebula
and subsequent gravitational instability of the dust layer results in
planetesimal formation.
The dust grain sedimentation timescale may be only $10^3$--$10^5$ yrs
(e.g., Hayashi, Nakazawa, \& Nakagawa 1985) and the gravitational 
instability operates over a timescale that is comparable to the
orbital period (reviewed by Papaloizou \& Lin 1995).
First-born planetesimals
have masses of a few $\times 10^{22}(a/40{\rm AU})^{3/2}$ g
(e.g. Hayashi, Nakazawa, \& Nakagawa 1985),
which is already comparable to the masses of the present EKB objects. 
However, nebula turbulence may prevent dust grains from settling
onto the equatorial plane, so that the gravitational instability 
does not occur (e.g. Weidenschilling \& Cuzzi 1993).
If this is the case, planetesimal accretion up to the present size of 
the EKB objects would require $10^8$--$10^9$ yrs
(e.g. Stern \& Colwell 1997), so that assumption 3 may be too
restrictive for our model of repeated encounters in an eccentric binary
and only the model of flyby stellar encounters (before dissolution
of a stellar aggregate) would be allowed.

A series of numerical simulations to test the effect of stellar companion 
encounters in protoplanetary disks has been performed. We consider
collisionless particles (corresponding to planetesimals), orbiting initially
on coplanar circles around a primary star (the proto-sun).
This particle disk encounters a hypothetical companion star.   
The orbital changes of the test particles are integrated taking into
account the gravitational forces of the primary and
the companion star using a fourth order predictor-corrector scheme.
Many different encounter geometries and companion masses have
been examined. If the scale length is defined by the pericenter
distance $q$ of the encounter, each encounter is characterized by:
the companion mass ($M_c$), the inclination angle of the companion orbit
relative to the initial disk ($\theta_c$), and the orbital energy or
eccentricity of the perturber ($e_c = 1$) (Ostriker 1994).

In the models, typically $10^4$ test particles were initially distributed
in the region {$a/q = 0.05$} -- {$0.8$}, where $a$ denotes
semimajor axis. The initial surface number density $n_{s0}$
is proportional to $a^{-1.5}$.
Since we consider test particles which do not interact with
one another, the particular choice of disk mass or 
surface number density profile
does not affect the generality of the results.
The initial eccentricity and inclination ($e_0$ and $i_0$)
of the particles are taken to be $\siml 0.01$.
Figure 2 shows the eccentricity and inclination of the particles
after the encounter as a function of $a/q$, in the case with
$e_0=i_0=0$ and $M_c=M_p$.
Inclination angle $\theta_c$ is (a) 5 degrees, (b) 30 degrees, 
and (c) 150 degrees 
with the line of nodes along the $x$-axis.
The spatial distribution in the case of $\theta_c = 30$ degrees
is shown in Fig.3.

As shown in Fig.2 the encounter leads to a strong increase in
$e$ and $i$ in the outer parts of the disk.
In the case of $\theta_c = 30$ degrees, 
$e$ and $i$ are pumped up only slightly
($\siml 0.01$) at $a/q \siml 0.2$, 
while they are pumped up
highly ($\simg 0.1$) at $a/q \simg 0.25$.
Note that the former condition is similar to the orbital stability 
condition of bounded three-body systems (e.g. Black 1982).
At $a/q \simg 0.3$ a large fraction of particles are ejected:
respectively 70\% and 95\% of objects with initial
$a/q \sim 0.5$ and $0.7$.
The remaining particles at $a/q \simg 0.3$ have large
eccentricities. The other $\theta_c$ cases show similar features.
The spike in $i$ at $a/q \sim 0.3$ coincides with the 3:1 commensurability
of the unperturbed disk orbital frequency and the companion's orbital frequency
at pericenter. Since the companion's angular velocity at pericenter is
$[2G(M_p+M_c)q^{-3}]^{1/2}$, the 3:1 commensurability is located at 
$a/q = [2(1+M_p/M_c)\times 3^2]^{-1/3} \simeq 0.30$, for prograde encounters.
Thus the highly localized character of the disk response in this
region appears to be associated with a corotation resonance occurring
near pericenter (Korycansky \& Papaloizou 1995).

As shown in Fig.3, long-lived features in the spatial distribution 
are the inclined bar-like envelope and prominent one-armed spiral, 
both due to close correlations in the longitudes of perihelion
and ascending node. Precession of the longitudes due to, for example,
the nebula potential would gradually destroy the features.

If the velocity dispersion is greater than the surface escape velocity
of a planetesimal, collisions between planetesimals 
is too destructive (Backman, Dasgupta, \& Stencel, 1995) and
growth of planetesimals is halted.
The surface escape velocity of a planetesimal with mass $m$ and
density $\rho$ is
$0.5 \times 10^5 (m/10^{24}{\rm g})^{1/3}(\rho/1 {\rm gcm}^{-3})^{1/6}$cm/s.
The velocity dispersion is 
$\sim (e^2+i^2)^{1/2} v_{\rm Kep} \simeq 0.5 \times 10^6 (e^2+i^2)^{1/2}
(a/40{\rm AU})^{-1/2}(M_p/M_{\odot})^{1/2}$cm/s, where $v_{\rm Kep}$ is
Keplerian velocity.
As a result, in the region where pumped-up $e$ or $i \simg 0.1$,
planetesimal growth would be inhibited.
The steep radial gradient of $e$ and $i$ seen in Fig.2 indicates 
that there exists a well defined boundary for planetesimal
growth at $a \sim 0.2$--$0.3q$: outside this region 
planetesimal growth is greatly inhibited while it is not affected at all
inside.

For different encounter parameters the distribution of the 
pumped-up $e$ and $i$ is generally very similar to Fig. 2, except for the
length scale $q$. In other words, for different encounters
the distribution of particles in Fig. 2 shifts towards larger or
smaller values of $a/q$, except that $i$ is not pumped in the special
case of a coplanar encounter. In general more massive companions and
lower inclination encounters yield stronger interactions. 
For example, the distribution shifts as $a/q \propto (M_c/M_p)^{-(0.2-0.25)}$.
Higher energy (i.e. more eccentric) encounters result in stronger effects
further into the disk owing to the improved coupling there.
Encounters with $\theta_c$ closer to 90 degrees
result in higher $i$ relative to $e$.
The $\theta_c = 150$ degree encounter has the same amplitude
of inclined angle as that with $\theta_c = 30$ degrees.
Hence, the pumped-up $i$ is similar, but $e$ is smaller (see Fig. 2),
because the parameter $\theta_c = 150$ degrees gives a retrograde encounter
and the relative velocity between disk particles and the
passing star is therefore significantly larger, resulting in only
very weak coupling.

As mentioned above, if the proto-sun had a transient binary companion,
the proto-sun may have experienced a few close
encounters with the companion before the binary system
broke up. In this case,
the individual encounters would have similar parameters and
$e$ and $i$ would be pumped up cumulatively with each encounter,
so that the perturbed forms of $e$ and $i$ would be preserved except
for shifts towards smaller values of $a/q$.

We shall now consider an encounter that gives
the required $e$ and $i$ distributions for the inner EKB.
As stated above, such encounters with $q$ on the order of $100$ AU
may be reasonable for the protosolar system.
We performed a similar simulation to that presented in
Fig. 2b ($\theta_c = 30$ degrees) 
except $\langle e_0^2 \rangle^{1/2}=\langle i_0^2 \rangle^{1/2}=0.01$.
Overall features of the pumped-up $e$ and $i$ are 
quite similar to Fig. 2b (see Fig. 4a).
With $q=160$AU, we randomly selected $500$ particles 
in the range $30$AU$<a<65$AU from the results,
to compare them with the observed numbers of EKB objects.
The selected distribution is shown in Fig. 4a.
For $a \simg 42$AU, $e$ and $i$ are as large as those of the observed
EKB objects.
Some objects that originally had larger $a$ are scattered to
this region with very high $e$ and $i$. 
However, at $a \siml 39$AU, $e$ is still small enough ($\siml 0.05$)
to allow the formation of a gap inside $39$AU via resonance sweeping,
without the need for any other processes,
e.g. long-term orbital destabilization.

In order to study the sweeping mean motion resonances we also
performed simulations similar to other authors 
(Malhotra 1995, Ida et al. 1999),
starting from the resultant distribution of particles after
the stellar encounter (Fig. 4a). The proto-Neptune with a mass of
$10^{29}$g (comparable to the present Neptunian mass)
was artificially moved from $23$AU to $30$AU (therefore the 3:2 resonance
moved from 30AU to 39.5AU),
on a circular zero inclination orbit.
We assumed a time dependence for 
the semimajor axis evolution given by:
$30\times [1 - (7/30)\exp(-t/5 \times 10^5{\rm yrs})]$AU, and
a migration timescale $a/\dot{a} = 2 \times 10^{6}$ yrs.
If we chose a longer migration time, 
more particles are captured by the 2:1 resonance 
and a gap is created interior to the resonance
while the capture probability 
of the 3:2 resonance would remain much as before (Ida et al. 1999).

The result after the sweeping is shown in Fig.4b.
The objects between $40$AU and $42$AU would be destabilized by
a long-term secular resonance (Duncan, Levison, \& Budd 1995).
The objects that have high eccentricity and are not
trapped by mean-motion resonances may experience
close encounters with Neptune and go to the 'scattered disk'.
Sweeping secular resonances, which we do not include in our simulations,
may alter the inclination distribution both near the 3:2 resonance
and beyond 42 AU (Malhotra, Duncan, \& Levison 1999).
Thus, our result is consistent with the observed distribution in Fig.1.
In particular, the puzzling high values of $e$
at $a \simg 42$AU are
explained without diminishing the capture probability of
the sweeping 3:2 resonance.
Different geometry of a stellar encounter, multiple encounters, or
an encounter with a passing binary system might result in a better
match.

The typical damping time of $e$ and $i$ due to hydrodynamic
gas drag at $40$AU is $10^9(m/10^{22}{\rm g})^{1/3}(e/0.1)^{-1}$ yrs 
(Adachi, Hayashi, \& Nakazawa 1976), for a typical minimum mass 
solar nebular model (Hayashi 1981).
This is much longer than the lifetime of disk gas, inferred
from observations, being of order $10^6$--$10^7$ yrs (e.g. 
Zuckerman, Forveille, \& Kastner 1995). 
Also the two-body relaxation time and
collision time for the presently estimated surface density at
$40$AU is longer than the Solar System age (Stern 1995, 1996;
Davis \& Farinella 1996).
Hence, the orbital elements of the present EKB objects should not have
changed significantly after the orbital perturbation. 
It is expected that the orbital distribution in $e$, $i$ and $a$
after the encounter reflects that observed today.

\section{Discussion}

Our simulations show that early stellar encounters would lead to 
interesting features in the young solar nebula that might explain
the structure of the outer part of the EKB.
The stellar encounters would occur
on timescales of dissolution of stellar aggregates, which
is of the order of $\sim 10^8$ yrs.
This may allow the EKB objects to grow to their observed sizes
before the encounters.
The objects initially inside 30AU would be strongly scattered to form
the 'scattered disk' during Neptune's migration (Duncan \& Levison 1997).
The objects with initial $a$ from 30AU to 40AU would be captured
by the sweeping of the 3:2 resonance with resultant high $e$ and $i$.
Outside 40AU, the stellar perturbations are strong enough to
pump up $e$ and $i$ to $\simg 0.1$.
Once their velocity dispersion is pumped up to
more than the surface escape velocity,
collisions between the EKB objects would produce copious amounts
of dust particles which would be removed by gas drag, Poynting-Robertson drag, 
and radiation pressure-driven ejection (Stern 1995; 
Backman, Dasgupta, \& Stencel 1995). The initial surface density is therefore 
eroded by virtue of its dynamical state, the present EKB objects
being remnants that have avoided significant erosion
(Stern 1996; Davis \& Farinella 1996). 
This result could explain the fact that the observationally inferred 
surface density in the EKB is much
lower than that extrapolated from a minimum mass solar nebula
model (e.g. Stern 1995; Weissman and Levison 1997).  
Detailed numerical modeling
of the subsequent collisional evolution of the perturbed EKB is required to
test this hypothesis.

Our model predicts that there should be a steep increase in
$e$ and $i$ with semimajor axis. In contrast, stirring by
by Earth-sized bodies would predict decrease in $e$ and $i$ 
and that by partial trapping by the Neptunian
sweeping 2:1 resonance predicts a 'cold' disk beyond 50AU.
Future observations can validate these
models by the trend of radial dependences in $e$ and $i$.
If $e$ and $i$ systematically increase beyond 50AU, our model
is supported.

The high eccentricities and inclinations that follow immediately
from such an encounter also have a number of consequences for extrasolar
planetary systems. Firstly, as stated previously, the augmented
velocity dispersion amongst planetesimals promotes the production of
dust particles. This can significantly increase the dust replenishment
rates and lead to more prominent circumstellar disks
around some main sequence stars 
(Stern 1995; Kalas \& Jewitt 1996; Holland et al. 1998).
The existence of the dust disks may reflect
stellar encounters in the formation epoch. Secondly,
as stated above, planetesimal growth could be forestalled in the outer
region of the disk by a stellar encounter.
This situation could be reflected in the fact that Neptune marks
the outer boundary of our planetary system at $30$AU. Thus the existence
of substantial planetary bodies outside $50$AU would be inconsistent
with our model.
Finally, we comment that recent advances in star formation theory and
observation suggest that such stellar encounters with disks as those
considered here should not be viewed as unique catastrophic events
but as an integral part of the star- and planetary-system formation process.

We thank the anonymous referee for helpful comments and
useful suggestions. SI acknowledges the hospitality of the
MPIA during his stay. JL is grateful to Dr. P. Kalas for comments
on dust disks.  

\newpage

\begin{reference}

\reference Adachi, I., Hayashi, C., \& Nakazawa, K. 1976. Prog. Theor. Phys.,
 56, 1756
\reference Backman, D. E., Dasgupta, A., \& Stencel, R. E. 1995, ApJ, 
 450, L35
\reference Black, D. C. 1982, AJ, 87, 1333 
\reference Davis, D. R. \& Farinella, A. P. 1996, Icarus, 125, 50 
\reference Davis, D. R. \& Farinella, A. P. 1998, LPSC abstract
\reference Duncan, M. J., Levison, H. F., \& Budd, S. M. 1995, \aj, 110, 3073 
\reference Ghez, A.M., McCarthy, D.W., Patience, J.L., \& Beck, T.L 1997
ApJ, 481, 378
\reference Gladman, B. et al. 1998, AJ, 116, 2042
\reference Goldreich, P. \& Ward, W. R. 1973, ApJ, 183, 1051
\reference Hahn, M. J. \& Malhotra, R. 1999, \aj, in press
\reference Hayashi, C. 1981, Prog. Theor. Phys. Suppl., 70, 35
\reference Hayashi, C., Nakazawa, K., \& Nakagawa, Y. 1985 
in Protostars and Planets II. eds. D. C. Black \& M. S. Matthew
(Tucson: Univ. of Arizona Press), 1100
\reference Hillenbrand, L.A. 1997, AJ, 113, 1733
\reference Holland, W., Greaves, J., Zuckerman, B., Webb, R.,
 McCarthy, C., Coulson, I., Walther, D., Dent, W., Gear, W., \& Robson,
 I. 1998, Nature, 392, 788
\reference Ida, S., Bryden, G., Lin, D. N. C., \&
 Tanaka, H. 1999, ApJ, submitted
\reference Jewitt, D., Luu, J. \& Trujillo, C. 1998, AJ, 115, 2125
\reference Kalas, P. \& Jewitt, D. 1996, AJ, 111, 1374
\reference Kroupa, P. 1995, MNRAS, 277, 1507
\reference Kroupa, P. 1998, MNRAS, 298, 231 
\reference K\"ohler, R \& Leinert, C.H. 1998 A\&A, 331, 977
\reference Laughlin, G. \& Adams, F. C. 1998, ApJ, 508, L171 
\reference Malhotra, R. 1995, \aj, 110, 420
\reference Malhotra, R., Duncan, M. J., \& Levison, H. F. 1999,
in Protostars and Planets IV eds. V. Mannings \& A. Boss
(Tucson: Univ. of Arizona Press), in press 
\reference Morbidelli, A. \& Valsecchi, G. B. 1997, Icarus, 128, 464
\reference Ostriker, E.C. 1994, ApJ, 424, 292
\reference Papaloizou, J.C.B. \& Lin D.N.C. 1995, ARA\&A, 33, 505
\reference Petit, J-M, Morbidelli, A. \& Valsecchi, G. B. 1999, Icarus, 
in press
\reference Safronov, V. S. 1969, Evolution of the Protoplanetary Cloud and
 Formation of the Earth and Planets. (Moscow: Nauka Press)
\reference Stern, S. A. 1991, Icarus, 90, 271
\reference Stern, S. A. 1995, \aj, 110, 856
\reference Stern, S. A. 1996, \aj, 112, 1203 
\reference Stern, S. A. \& Colwell, J. E. 1997, ApJ, 490, 879
\reference Weidenschilling, S. J. \& Cuzzi, J. N. 1993, 
 in Protostars and Planets III eds. E.H. Levy \& J.I. Lunine 
 (Tucson: Univ. of Arizona Press), p. 1031
\reference Zuckerman, B., Forveille, \& Kastner, J. H. 1995, Nature, 373, 494

\end{reference}

\begin{figure}
\caption{Orbital distribution 
(eccentricity, inclination, and semimajor axis $a$) 
of the observed EKB objects (from Marsden's web site,
http://cfa-www.harvard.edu/$^\sim$graff/lists/TNOs.html).
The objects observed at multiple oppositions or
during a relatively long period are plotted.
Inclination is plotted in radians [$0,2\pi$].
The unit of $a$, AU, is the mean distance between the sun and Earth.
The present locations of Neptune and its 3:2 and 2:1 resonances are 30AU,
39.5AU, and 47.8 AU, respectively.}
\end{figure}

\begin{figure}
\caption{
Orbital eccentricity $e$ and inclination $i$ of the particles
pumped up by a companion encounter,
as a function of semimajor axis $a$.
(Initial $e$ and $i$ are zero.)
Semimajor axis is scaled by the pericenter distance $q$.
Inclination is given in radians [$0,2\pi$].
$M_c = M_p$ and parabolic encounter are assumed for the companion:
(a) the case with $\theta_c  =5$ degrees,
(b) the case with $\theta_c  =30$ degrees,
(c) the case with $\theta_c =150$ degrees.}
\end{figure}

\begin{figure}
\caption{
The spatial distribution of particles after a companion encounter,
projected into the initial orbital plane ($xy$-plane) and 
projections perpendicular to the $x$ and $y$-axis,
20 rotation times (at $a/q=1$) after
the encounter.
The results of the encounter in Fig.2b ($\theta_c = 30$ degrees) is shown.
The $x$, $y$, and $z$-axes are scaled by the pericenter distance $q$
of the companion encounter.
The companion enters from the lower-right, passes through pericenter
$(x/q,y/q,z/q)=(0,1,0)$, and exits to lower-left.
} 
\end{figure}

\begin{figure}
\caption{ Orbital distribution 
of the predicted EKB objects:
(a) predicted distribution after the stellar companion encounter
(before the sweeping resonances),
(b) prediction by combined simulations
of the stellar companion encounters and the sweeping mean motion
resonances associated with Neptune migration.
For parameters of the companion encounters and Neptune's migration, 
see text.  }
\end{figure}

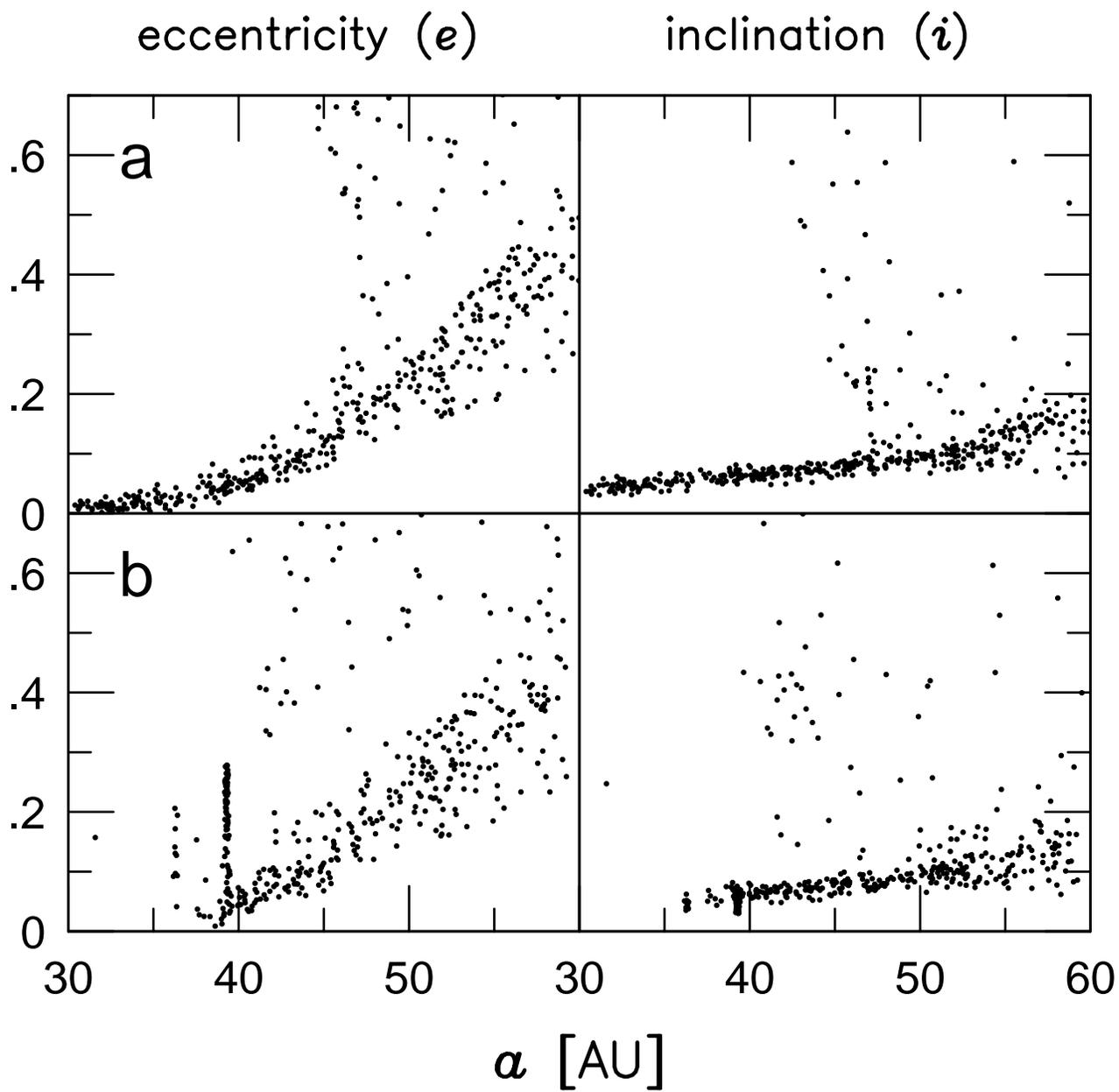
\end{document}